\documentclass[pra,twocolumn,showpacs,preprintnumbers,amsmath,amssymb,superscriptaddress]{revtex4-1}

\usepackage{graphicx}

\begin{document}


\title{Imprinting Light Phase on Matter Wave Gratings in Superradiance Scattering }

\author{Xiaoji Zhou}
\email[E-mail: ]{xjzhou@pku.edu.cn}
\author{Fan Yang}
\author{Xuguang Yue}
\author{T. Vogt}
\author{Xuzong Chen}
\email[E-mail: ]{xuzongchen@pku.edu.cn}
\affiliation{ School of Electronics Engineering $\&$ Computer
Science, Peking University, Beijing 100871, P. R. China}

\date{\today}

\begin{abstract}

Superradiance scattering from a Bose-Einstein condensate is studied
with a two-frequency pumping beam. We demonstrate the possibility of
fully tuning the backward mode population as a function of the
locked initial relative phase between the two frequency components
of the pumping beam. This result comes from an imprinting of this
initial relative phase on two matter wave gratings, formed by the
forward mode or backward mode condensate plus the condensate at
rest, so that cooperative scattering is affected. A numerical
simulation using a semiclassical model agrees with our observations.
\end{abstract}

.

\maketitle


Superradiance from a Bose-Einstein Condensation (BEC) offers the possibility
of studying a novel physics associated with cooperative scattering of light
in ultracold atomic systems.  A series of experiments~\cite{Inouye1999science,
Schneble2003scince, 1999, Yutaka} and related theories~\cite{Moore1999prl,
Zobay2006pra, Pu2003prl} have sparked new interests in matter wave
amplification~\cite{Inouye1999science, 1999}, holographic
storage~\cite{Yutaka}, scattering spectroscopy in optical lattice~\cite{xu},
coherent imaging~\cite{sadler}, coherent atomic recoil
lasing~\cite{Courteille1, Courteille2,zhou}, and quantum states storage and
retrieval~\cite{Lukin,Matsuk}.

In a typical BEC superradiance experiment, an elongated condensate
is illuminated  by an off-resonant pumping laser pulse along its
short axis. Due to the phase-matching condition and mode
competition, highly directional light is emitted along the long axis
of the condensate, in the so-called end-fire modes. Consequently,
the recoiled atoms acquire a well-defined momentum at $\pm
45^{\circ}$ angles with respect to the pumping laser direction.
These atomic modes are referred to as forward modes. This forward
scattering is interpreted as optical diffraction from a matter wave
grating~\cite{Inouye1999science, Moore1999prl}. Meanwhile, atoms in
the condensate may scatter photons in the end-fire modes back into
the pumping mode and recoil at $\pm135^{\circ}$ angles, forming the
so-called backward modes~\cite{Schneble2003scince, Pu2003prl}, when
the pumping pulse is short and intense. This pattern was interpreted
as a result of diffraction of atoms off a light grating. A four wave
mixing interpretation was proposed, involving two optical
fields--the pumping laser field and an end-fire mode, and two matter
wave modes--the condensate and a mode of momentum~\cite{ Pu2003prl}.
\begin{figure}
   \begin{center}
    \includegraphics[width=8cm]{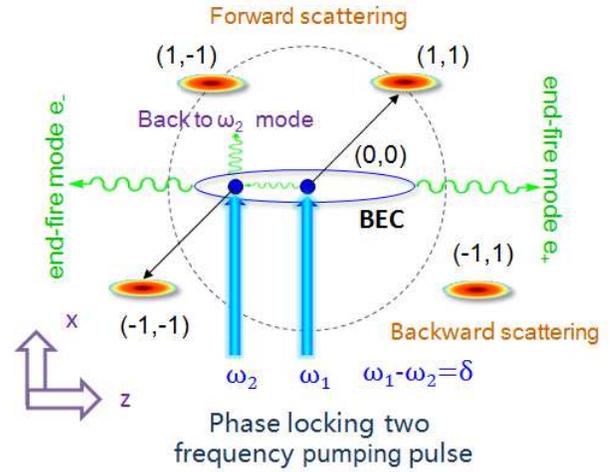}
   \end{center}
   \caption{(Color online)  Schematic
   diagram of our experiment. The two-frequency pumping beam is incident along the short $x$ direction, with a linear polarization along the $y$ direction. The atomic side modes are denoted in momentum space, each labeled with a pair of integers which describe the order in the $x$ and $z$ directions, respectively. Within this notation, atoms of the condensate at rest are in mode $(0,0)$. A forward scattering event transfers an atom from mode $(n,m)$ to mode $(n+1,m\pm 1)$, and a backward event transfers one to mode $(n-1,m\pm 1)$.The end fire mode $e_{\pm}$ is along the long axis of the condensate in $z$ direction.}
   \label{setup1}
\end{figure}

There is an energy mismatch of four times the recoil frequency for
this backward scattering, due to the increased kinetic energy of
recoiled atoms~\cite{Schneble2003scince, Zobay2006pra}. Recently, a
two-frequency-pumping scheme has been
implemented~\cite{KMRvdStam2007arxiv, Bar-Gill2007arxiv, yang},
where the pumping beam consisted of two frequency components and the
frequency difference was controlled to compensate for the energy
mismatch and excite the backward scattering on a long time scale
with a weak pump intensity. The presence of the backward mode in the
spectroscopic response~\cite{Bar-Gill2007arxiv} and the enhancement
of the diagonal sequential scattering~\cite{yang} have been
reported. In those experiments the relative phase between the two
pumping frequency components is maintained constant. Although phase
is a very important factor for understanding interference and
coherence phenomena, phase effects on matter wave gratings and
cooperative scattering have not been reported so far.

We present here our new experimental results that show a possibility
to fully control the backward mode population as a function of the
locked initial relative phase between the two pumping frequencies. A
theoretical analysis of the cooperative process is used to confirm
and explain this phase dependence. It requires to consider the
relationship between matter wave gratings and optical waves beating
for the first order scattering. The initial relative phase of the
pumping beam is imprinted into two matter wave gratings, one formed
by the condensate with a forward mode, the other formed by the
condensate with a backward mode.

\begin{figure}
  \begin{center}
        \includegraphics[width=8cm]{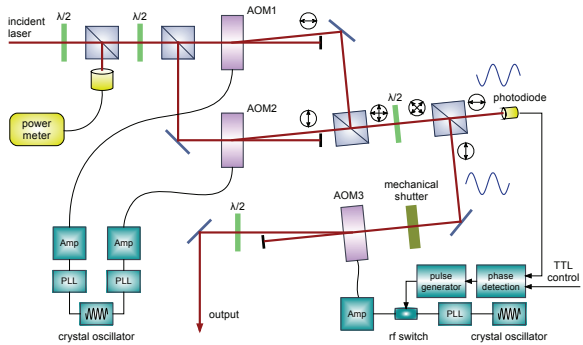}
   \end{center}
   \caption{Two-frequency superradiance scattering experimental setup. The pump laser beam is split into two beams with orthogonal polarizations using a PBS. With two AOMs, each beam frequency is shifted about 90 MHz, with a little difference of $\Delta\omega$. In order to keep the coherence of the two beams after frequency shifting, the phase-locked-loops of the radio frequency (rf) sources for the two AOMs are locked to the same crystal oscillator. A phase detection circuit is used to detect the initial relative phase of beating signal of the two frequency beams after they are combined. Note that the beat signal of the pump pulse is opposite in phase with the signal measured by the photodiode.}
   \label{twofrequency}
\end{figure}

In our experiment, a nearly pure Bose-Einstein condensate of about $2\times10^{5}$ $^{87}$Rb atoms in the $|F=2, m_{F}=2\rangle$ hyperfine ground state is generated in a quadrupole-Ioffe-configuration magnetic trap, with Thomas-Fermi radii of $50 \mathrm{\mu m}$ and $5 \mathrm{\mu m}$ along the axial and radial directions~\cite{yang}. The pumping pulse with two-frequency components is incident along the short axis of the condensate, with its polarization perpendicular to the long axis, as shown in Fig.~\ref{setup1}. This arrangement of polarization induces Rayleigh superradiance where all side modes possess the same atomic internal state~\cite{Inouye1999science}. To get the two-frequency pumping beam, a laser beam from an external cavity diode laser is split into two equal-intensity beams. Their frequencies are shifted
individually by acousto-optical modulators (AOM1 and AOM2) which are driven by phase-locked radio frequency signals. Therefore the frequency difference $\Delta\omega$ between the two beams can be controlled precisely, as shown in Fig.~\ref{twofrequency}. After that, they are recombined to form our linear-polarized two-frequency pump beam, which is red detuned by $2\pi \times 1.5 \mathrm{GHz}$ from the $|F=2, m_{F}=2\rangle$ to $|F'=3, m_{F}'=3\rangle$
transition.

\begin{figure}[!htbp]%
   \begin{center}
    \begin{picture}(0,0)
      \put(-10,110){(a)}\quad
       \end{picture}
    \includegraphics[width=6 cm]{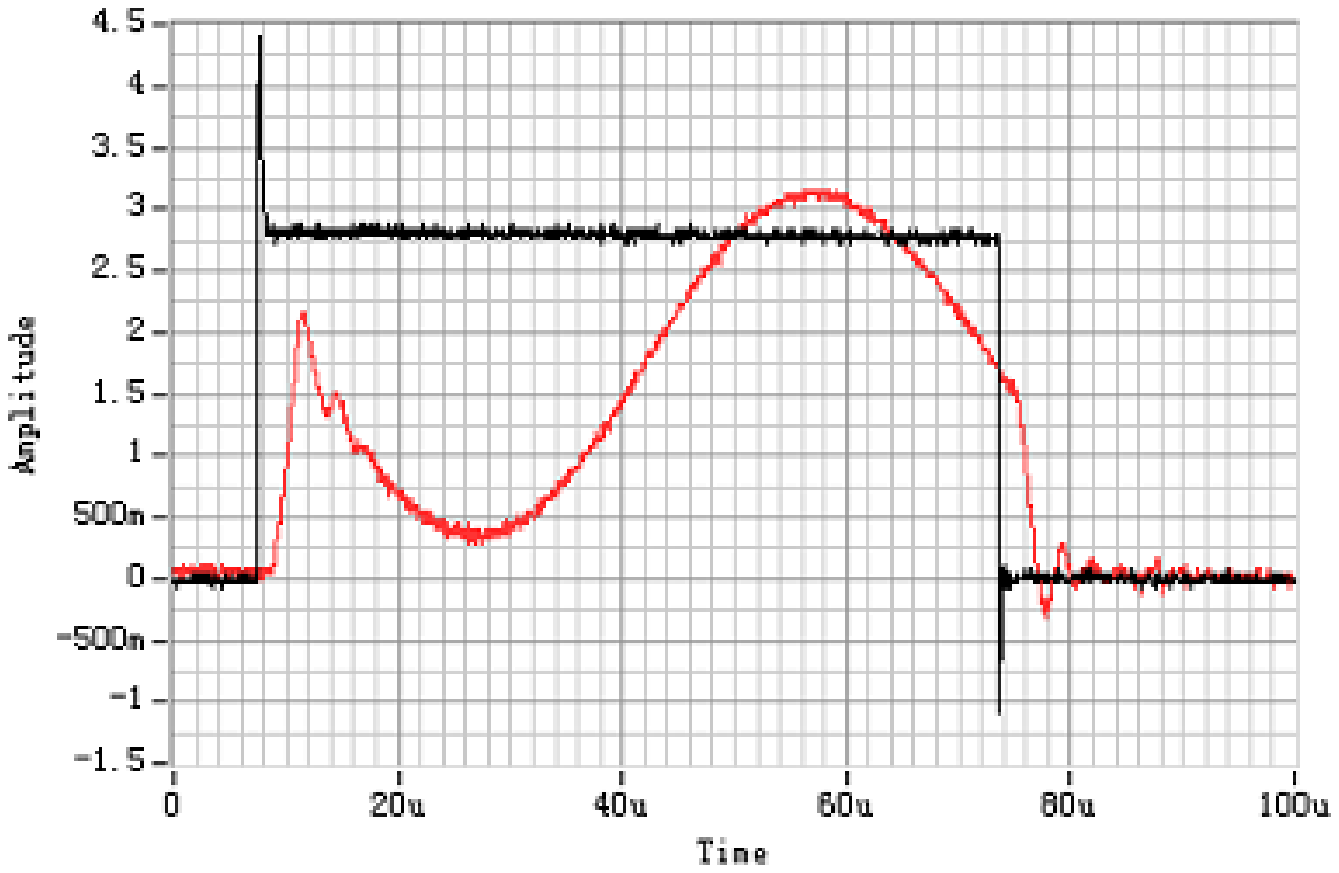}\quad\quad
    \begin{picture}(0,0)
      \put(-10,110){(b)}\quad
       \end{picture}
    \includegraphics[width=6 cm]{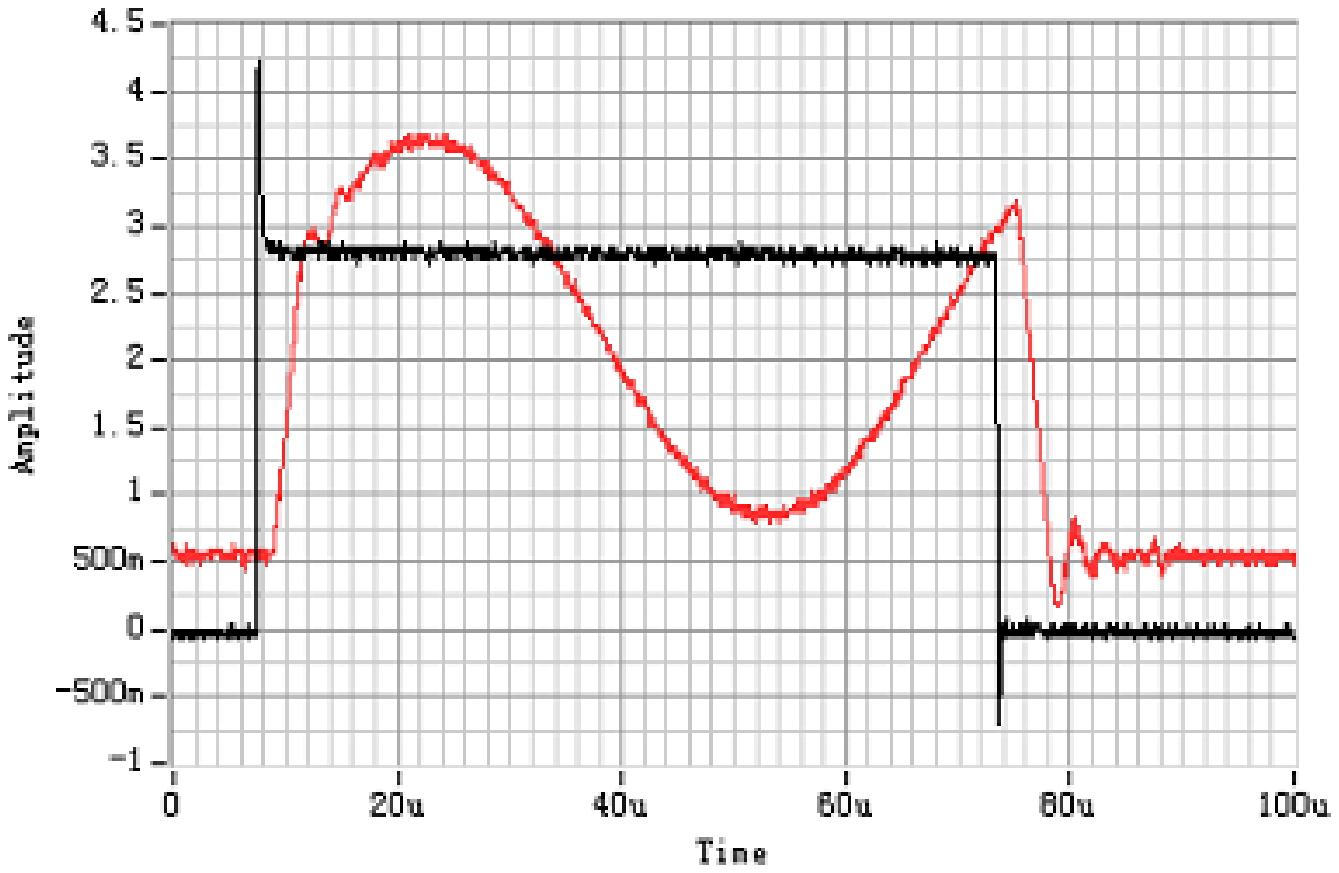}\\
     \end{center}
      \begin{center}
    \begin{picture}(0,0)
      \put(-10,100){(c)}\quad
    \end{picture}
    \includegraphics[width=6 cm]{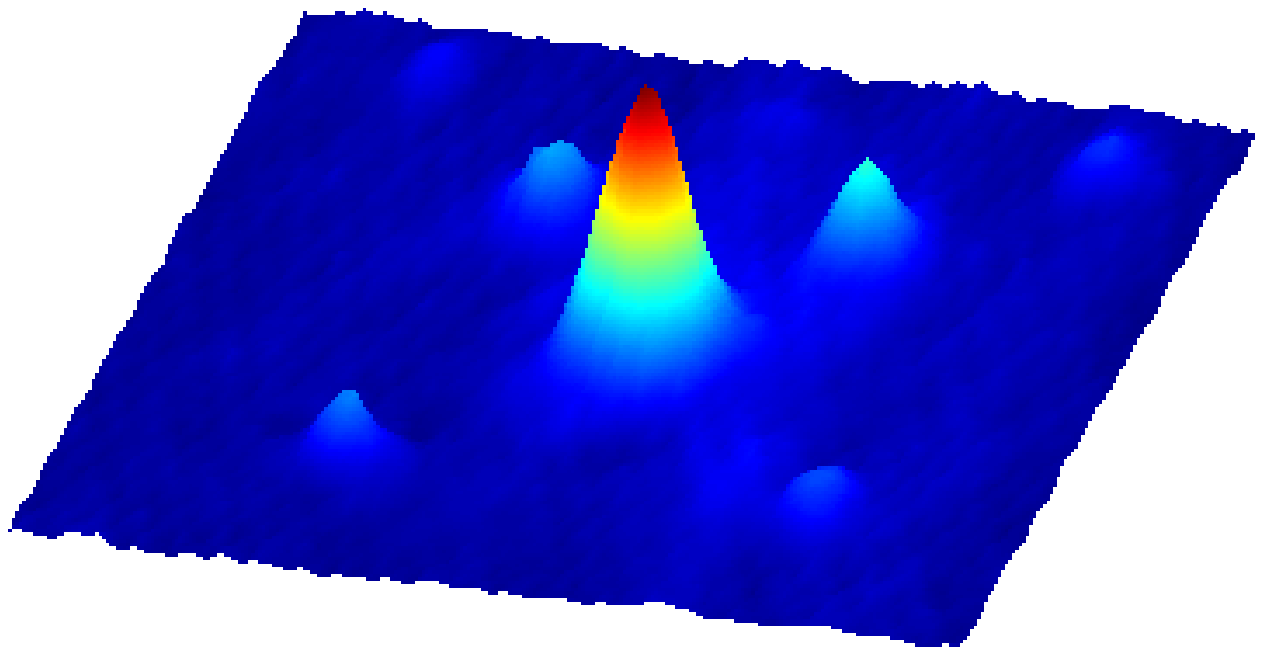}\quad\quad
    \begin{picture}(0,0)
      \put(-10,100){(d)}\quad
    \end{picture}
    \includegraphics[width=8 cm]{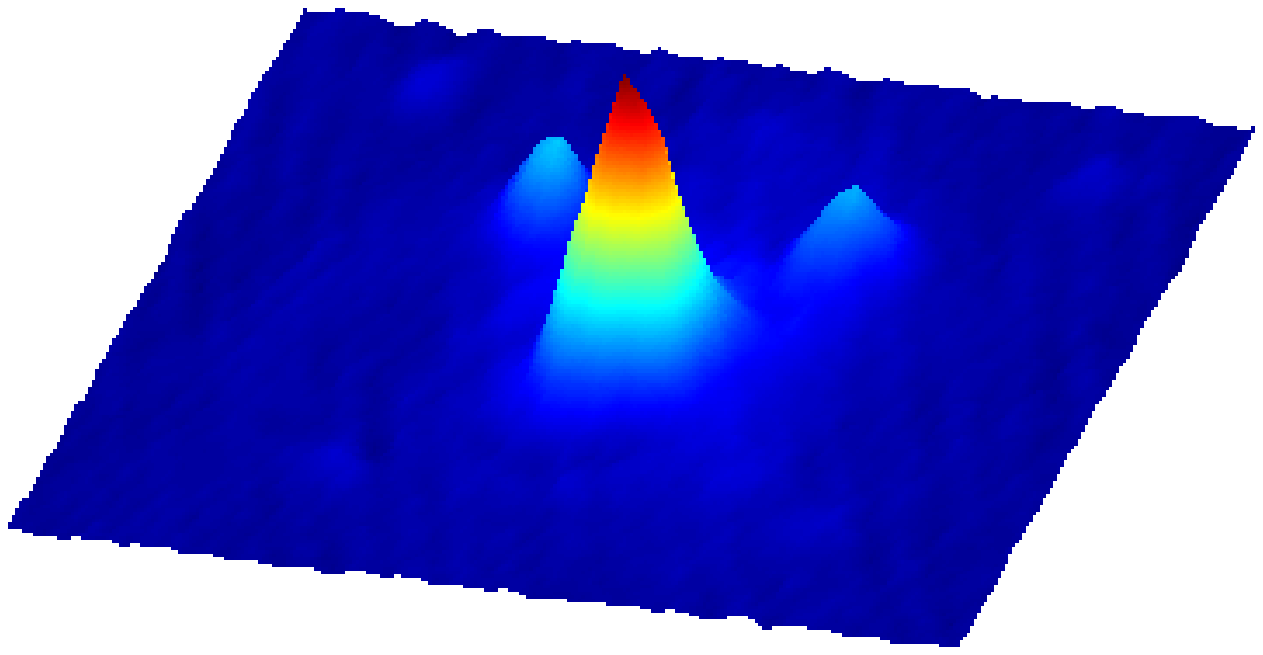}\\
    \begin{picture}(0,0)
      \put(-20,150){(e)}\quad
    \end{picture}
    \includegraphics[width=7 cm]{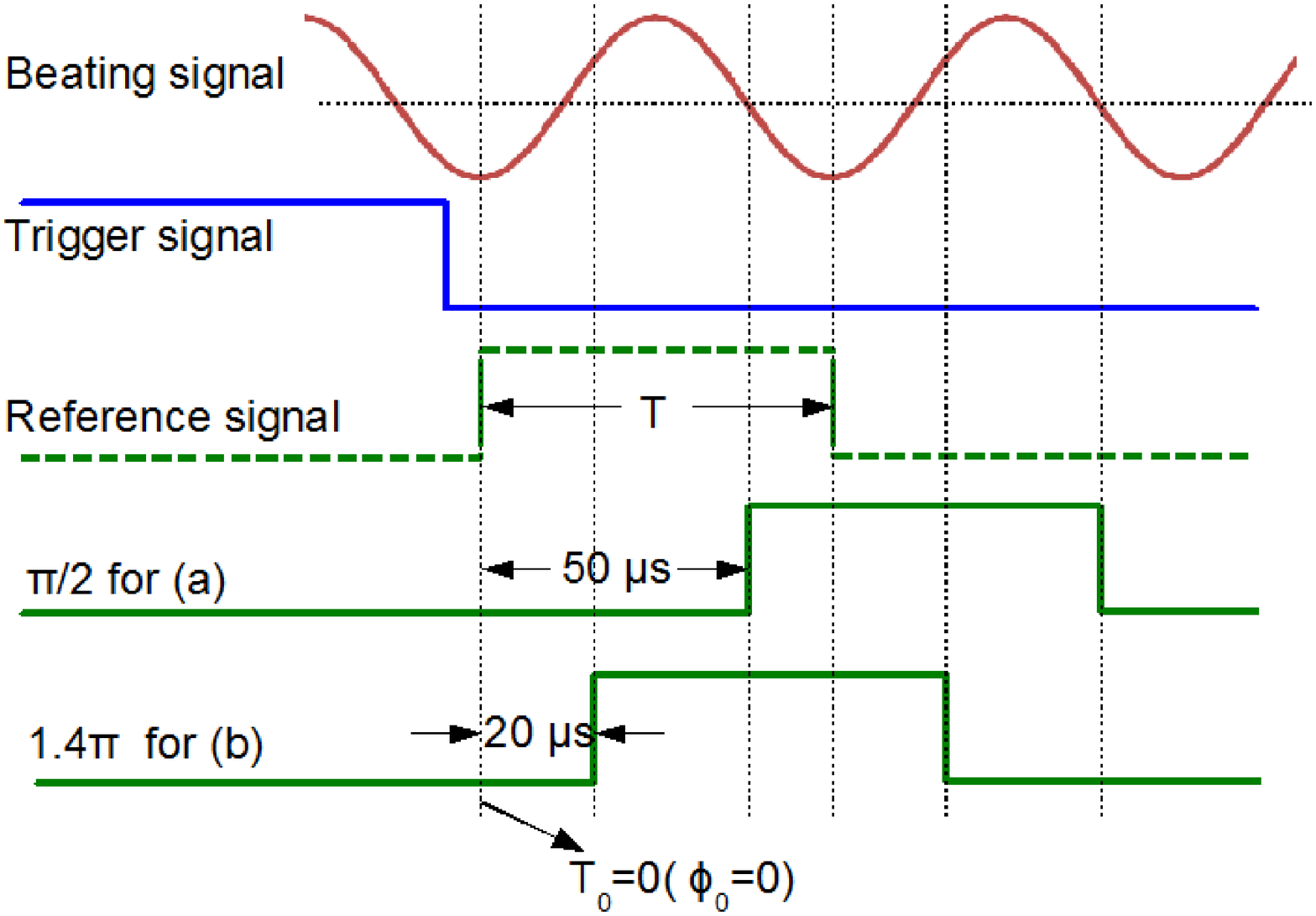}
   \end{center} 
   \caption{Superradiance with different initial relative phases and time control sequence.
   (a) and (b): Initial relative phase of $\pi/2$ and $1.4\pi$ corresponding to a generation time of 50$\mu$s and 20$\mu$s respectively.
   (c) and (d):  Superradiance patterns corresponding to (a) and
   (b), respectively.
    (e) Control sequence for generating the pump pulse. Note that there
     is a phase difference of $\pi$ between the phase shown in figures (a) and (b) and that of the real pumping pulse. }
   \label{intphase}
\end{figure}

The frequency difference $\Delta\omega$ is chosen to be $2\pi\times15$kHz (the
corresponding period $T$ is then $66.67 \mu$s), i.e. four times the recoil
frequency ($4\omega_{r} = 2 \hbar k_{l}^{2}/M$) so that we reach the two-photon resonance condition taking into account the recoil energies deficit for the backward scattering. The magnetic trap is shut off immediately after the pumping pulse, and the distribution of atomic side modes is measured by absorption imaging after $21\mathrm{ms}$ of ballistic expansion, as shown in
Fig.~\ref{intphase}.

The experiment is repeated with different initial relative phases
between the two frequency components. Monitoring the beating signal
$\cos(\Delta \omega t+\phi)$ on a photodiode, a light pulse of
duration $T$ with a definite initial relative phase $\phi_0=\Delta
\omega t_0+\phi$ can be generated. Different initial relative phases
$\phi_0$ varying between $0$ and $2\pi$, are obtained by switching
on the pulse at different time $t_0$, which we call ``generation
time", as shown in Fig. 3(e). The generation time $t_0$ and the
switch off time $t_0+T$ can be controlled precisely with an AOM.
Figure~3(a) and (b) demonstrate the initial relative phases of the
two frequency pumping beam being $\pi/2$ and $1.4\pi$, the
corresponding generation time $t_0$ being $50\mu$s and $20\mu$s,
respectively. This phase can be read out related to the reference
signal corresponding to $\phi_0=0$.

\begin{figure}
   \begin{center}
     \includegraphics[width=6cm]{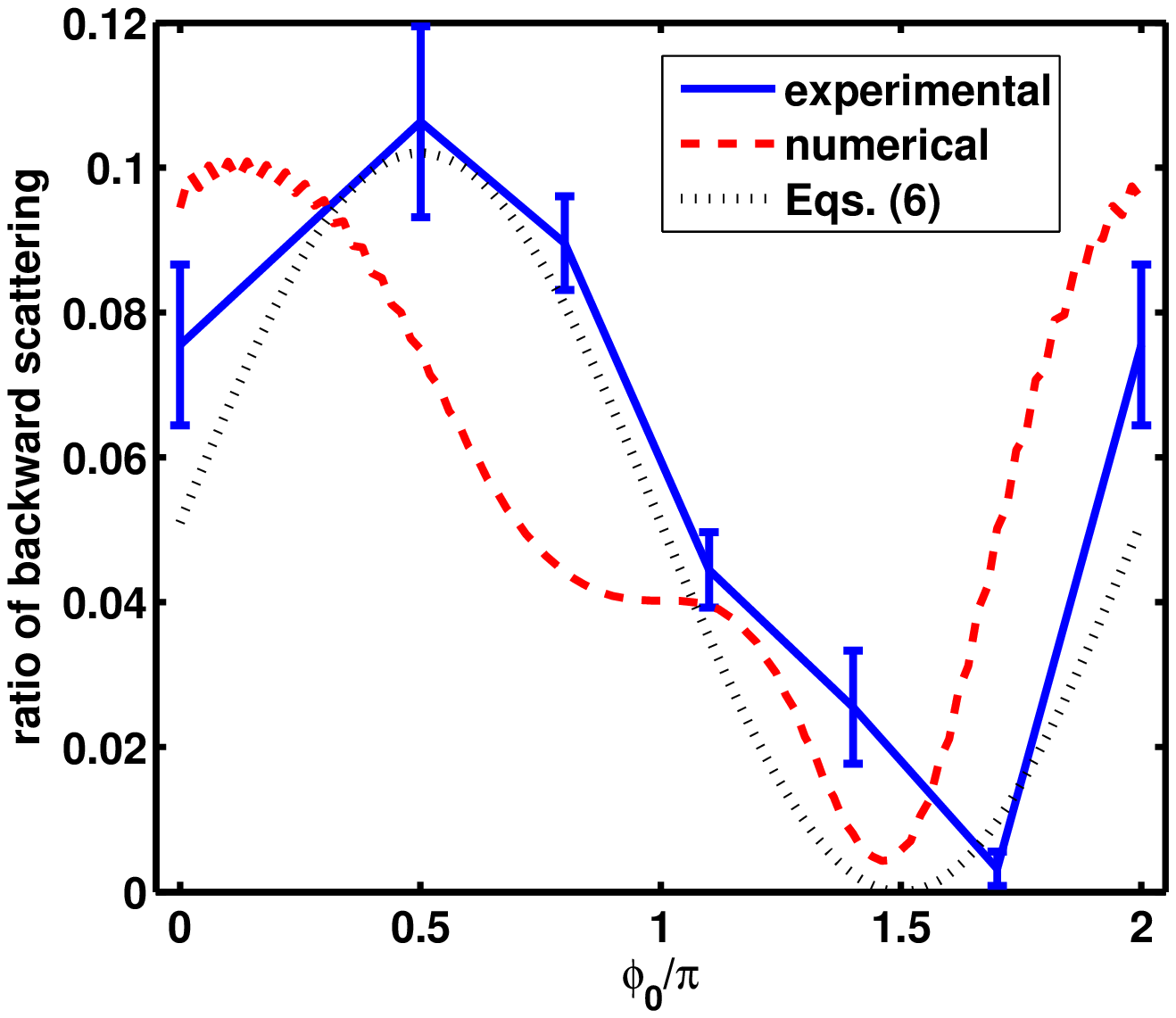}
   \end{center}
   \caption{(Color online) Experimental results and theoretical analysis about
   the ratio of backward scattered atoms to the total atom number vs the initial
   phase $\phi_{0}$ between the two pump beams.  Each point is an average on four
   experimental data. The dashed line is a numerical simulation using a
   semi-classical theory depicted in the text with one initial seed and c
   oupling factor $g=1.5\times 10^{6}$. The dotted line is drawn using
   Eq.(\ref{results}) with $\Delta\varphi_{0,1}=\pi/2$.}
  \label{experdata}
\end{figure}

In Fig.~\ref{intphase} (c), the backward mode is obvious for a
relative initial  phase $\phi_{0}=\pi/2$ while in
Fig.~\ref{intphase} (d), when the phase is $\phi_{0}=1.4\pi$,
backward scattering is almost suppressed. The ratio between the
backward population and the total atom number is plotted in
Fig.~\ref{experdata} versus different phases $\phi_0$, with a pulse
duration equal to $66.67 \mu$s or one cycle, a time step between two
points given by $\Delta t_{0}=10\mu$s and each point being the
average on four experimental data. The error bar is their standard
deviation. For each experiment, the atom number in the BEC, the
initial quantum noise and the temperature are not exactly the same.
These fluctuations are the main reason for the experimental
uncertainty. Nevertheless, the number of backward scattered atoms
shows a high sensitivity to the initial relative phase between the
two frequency components and it is possible to obtain an almost
complete cancellation of the backward scattering for
$\phi_{0}=3\pi/2$.

\bigskip

Such a result was  not expected considering in particular the
interpretations given in previous experiments. Especially, the
exponential growth model of scattered atoms used in previous
articles \cite{Inouye1999science} would not accurately account for
the phase dependence. Even considering the depletion of the original
mode of the condensate it writes ${\mathrm{d} N_s(t)}/{\mathrm{d}
t}= G(t) \left( N_0 - N_s(t) \right) N_s(t)$, where $N_S$ is the
number of scattered atoms and $N_0$ the initial number of atoms in
the  condensate~\cite{Inouye1999science}. The gain $G(t)$ is
proportional to the pump intensity $I(t)$. This logistic
differential equation can be solved analytically as $ N_s(t) =[N_0
e^{N_0 \int_0^t G(t) \mathrm{d}t}]/{[N_0 -1 + e^{N_0 \int_0^t G(t)
\mathrm{d}t}]}$. Within this model, the number of scattered atoms is
determined by the integration of the pump intensity $I(t)$, thus
proportional to $\int_{0}^{t}I(t)dt = I_0 \int_{0}^{t}
[1+\cos(\Delta \omega t +\phi_0)]dt\geq0$. When the pumping pulse
duration equals to an exact cycle, the integration of the pump
intensity is always the same, regardless of the relative initial
phase between the two frequency components. This means, within this
nonlinear growth model, the number of scattered atoms is independent
of the initial relative phase.

The above model is not valid since more than one scattered mode
should  be taken into account. Actually, the different modes form
matter wave gratings and affect the scattering behavior. To give a
physical explanation, we need to clarify the relationship between
the population of the different modes and the initial relative phase
$\phi_0$. Here we restrain our analysis to a set of diagonal  modes
: $(-1, -1)$, $(0, 0)$, and $(1, 1)$. Results are similar for the
other diagonal modes ($(1, -1)$, $(0, 0)$, and $(-1, 1)$) because of
symmetry. Using the slowly varying envelope approximation
(SVEA)~\cite{Zobay2006pra}, the population of mode $(-1, -1)$ is
given by its wave function $\psi_{-1,-1}(\xi,\tau)$ :
\begin{equation}\label{number}
\frac{\partial N_{-1,-1}(\tau)}{\partial
t}=\int_{-\infty}^{\infty}d\xi \left(\frac{\partial\psi_{-1,-1}}
{\partial t}\psi_{-1,-1}^{*} +\frac{\partial\psi_{-1,-1}^{*}}
{\partial t}\psi_{-1,-1}\right).
\end{equation}
We use the semiclassical Maxwell-Schr\"{o}dinger equations to
describe  the coupled dynamics of matter-wave and optical fields.
Although this theory can not be used to analyze the initial quantum
spontaneous process, the pulse in the experiment is about several
tens of $\mu s$ and far beyond its initial quantum characteristic.
So the effect of the pump beam phase on the initial seeds can be
omitted to analyze the experimental
data~\cite{Zobay2006pra,Bar-Gill2007arxiv,yang}. Considering the
trapped BEC is tightly constrained at its short axis degrees and the
Fresnel index number of the optical field is around 1, we consider a
one-dimensional semiclassical model including spatial propagation
effects. Then the envelope function of the end-fire mode optical
fields with a pump laser consisting of two components having same
intensity and polarization but different frequencies $\omega_{l}= c
k_{l}$ and $\omega_{l}-\Delta\omega$, is given
by~\cite{Zobay2006pra,yang}
\begin{eqnarray}\label{e}
    e_{-}(\xi,\tau) &=& -i \kappa(\phi_{0})  \sqrt{\frac{L}{k_{l}}}
    \int_{\xi}^{\infty}\mathrm{d}\xi'
    (\psi_{0,0} \psi_{1,1}^{*}
    \nonumber\\&+&
    \psi_{-1,-1} \psi_{0,0}^{*}e^{-i2\tau}),
\end{eqnarray}
where $\tau = 2\omega_{r}t$ and $\xi = k_{l}z$. $\kappa(\phi_{0})=g
\left(1+ e^{i (2 \tau + \phi_{0})}\right)$ is connected with the
initial relative phase $\phi_{0}$ and the coupling factor between
light and atom $g=\sqrt{3 \pi c R / (2 \omega_{l}^{2} A L)}$. $R$ is
the Rayleigh scattering rate of the pump component and $L$ the BEC
length. It indicates that the end-fire mode field $e_{-}$ is due to
the coherence between different modes, such as modes $(0,0)$ and
$(1,1)$, $(-1,-1)$ and $(0,0)$,  and spatial overlap between these
modes is needed. There exists $4\omega_{r}$ frequency difference
between adjacent modes.

The coupled evolution equations of atomic side modes is given by
\begin{equation}\label{eq1}
\frac{\partial\psi_{-1,-1}(\xi,\tau)}{\partial \tau}=- i
\kappa^{*}(\phi_{0}) e_{-} \psi_{0,0} e^{i 2\tau}
\end{equation}
This equation describes the atom exchange between modes $(-1,-1)$
and $(0,0)$ through the pump laser and end-fire mode field. An atom
in $(-1,-1)$ mode may absorb a laser photon and emit it into an
end-fire mode $e_{-}$, and the accompanying recoil drives the atom
into the $(0,0)$ mode. The atom may absorb an endfire mode photon
and deposit it into the laser mode, to form the backward scattering
mode $(-2,-2)$ which process is very weak and omitted. Here we have
omitted the dispersion term, spatial translation, photon exchange
between modes and non-diagonal modes connected with $e_{+}$ such as
mode $(-2,0)$ and $(0,-2)$~\cite{Zobay2006pra,yang}.

Inserting Eq.~(\ref{e}) and (\ref{eq1}) into Eq.~(\ref{number}), we
can get the evolution equation of $N_{-1,-1}$
\begin{eqnarray}{\label{a}}
 \frac{\partial N_{-1,-1}(\tau)}{\partial \tau} &=& -2\frac{g^2Lc}
 {\omega_rc}[1+\cos(2\tau +\phi_0)]
 \nonumber\\
 &\times&\left[C_{0,1}e^{i2\tau}+C_{-1,0}\right]+c.c.,
\end{eqnarray}
where the envelope function of each side mode $\psi_{m,m}$ is
written as $\psi_{m,m}=|\psi_{m,m}|e^{-i\varphi_{m,m}}$ with the
phase $\varphi_{m,m}$ assumed to be space independent. Here we
define $C_{m,n}(\tau)=\int_{-\infty}^{\infty}d\xi
\tilde{C}_{m,n}(\xi,\tau)$=$|C_{m,n}(\tau)|e^{-i\Delta\varphi_{m,n}}$
with the phase difference between the matter waves
$\Delta\varphi_{m,n}=\varphi_{m,m}-\varphi_{n,n}+\varphi_{0,0}-\varphi_{-1,-1}$,
and the interference grating
$\tilde{C}_{m,n}(\xi,\tau)=\psi_{0,0}\psi^*_{-1,-1}\int_\xi^\infty
d\xi^\prime\psi_{m,m}\psi^*_{n,n}$. Then we have
$\Delta\varphi_{0,1} =
2\varphi_{0,0}-\varphi_{1,1}-\varphi_{-1,-1}$, and
$\Delta\varphi_{-1,0} = 0$.

In this case the above Eq.~(\ref{a}) can be expanded using trigonometric formula
\begin{eqnarray}
  \frac{\partial N_{-1,-1}(\tau)}{\partial \tau}&=&-\frac{g^2L}{\omega_rc}
\left\{\left[|C_{0,1}|2\cos(2\tau-\Delta\varphi_{0,1})\right.\right.\nonumber\\
  &+&\cos(\phi_0+\Delta\varphi_{0,1})+\cos(4\tau+\phi_0-\Delta\varphi_{0,1})]\nonumber\\
  &+&\left.2|C_{-1,0}|[1+\cos(2\tau+\phi_0)]\right\}
\end{eqnarray}
In a first approximation, we replace $|C_{m,n}|$ by its time average
$\overline{|C_{m,n}|}$. Equation~(5) can then be integrated from $0$
to one cycle, and the cosine functions of time vanish, leaving only
the $\cos(\phi_0 +\Delta\varphi_{0,1})$ term. A simpler expression
for $N_{-1,-1}$ at the end of the pumping pulse($\Delta\omega
t=2\pi$ or $\tau=\pi$) is obtained:
\begin{equation}{\label{results}}
  N_{-1,-1}(\pi)=-2\pi\frac{g^2L}{\omega_rc}\overline{|C_{0,1}|}\cos(\phi_0
  +\Delta\varphi_{0,1})+\alpha,
\end{equation}
The first term of Eq.~(\ref{results}) indicates that the population
of side mode $(-1,-1)$ is connected with the module of four waves
$|C_{0,1}|= |\psi_{0,0}\psi^*_{-1,-1}|\int_\xi^\infty
d\xi^\prime|\psi_{0,0}\psi^*_{1,1}|$, the relative phase $\Delta
\varphi_{0,1}$ and the optical initial relative phase $\phi_{0}$,
the coefficient of the cosine function determining the amplitude of
the oscillation. The second part $\alpha$ approximated to be a
constant in the above expression should actually include other terms
due to the residual time dependent part of $|C_{m,n}|$ and may also
slightly depend on $\phi_0$ . This dependence may be neglected in a
first approximation as the time-dependent terms in  Eq. (5) have
different time-dependent signatures and may cancel each other after
integration. The main purpose of Eq.~(\ref{results}) is to show the
dependence of the backward scattering on the initial relative phase
$\phi_0$, although it is hard for us to prove its positivity due to
the complexity of $\alpha$. However, numerical calculations
demonstrate that this backward scattered atom number is always
positive.

To give a physical explanation to this initial relative phase
dependence and because of the assumption that the phase
$\varphi_{m,m}$ of each side mode does not depend on space, we can
focus our attention on the terms $\tilde{C}_{m,n}$  since the
spatial integral from $-\infty$ to $\infty$ in the definition of
$C_{m,n}$ does not change the relative phase of the matter waves.
The term $\tilde{C}_{0,1}=\psi_{0,0}\psi^*_{-1,-1}\int_\xi^\infty
d\xi^\prime \psi_{0,0}\psi^*_{1,1}$  indicates the presence of two
sequential diffractions by two gratings, which we analyze as
follows. With the spatial coordinates described in Fig.~1, the
incident laser light traveling in the $x$ direction is diffracted by
the grating formed by modes $(0,0)$ and $(1,1)$, which is along the
diagonal of the $x-z$ plane, and can be expressed as
$\int_{z}^\infty dz |\psi_{0,0}\psi_{1,1}^*
|\cos(kx+kz+\varphi_{1,1} -\varphi_{0,0})$. The scattered light
(end-fire mode light) propagating along the long axis of the
condensate in $-z$ direction is then diffracted again by the grating
formed by modes $(0,0)$ and $(-1,-1)$, which is also along the
diagonal of the $x-z$ plane, and can be described as
$|\psi_{0,0}\psi_{-1,-1}^*| \cos(kx+kz-\varphi_{-1,-1}
+\varphi_{0,0})$, resulting in a scattered light traveling back into
$x$ direction which is highly correlated to the backward scattered
atom number. The spatial integral in the former grating  describes
the propagating effect. Here a phase shift of $\pi/2$ takes place as
the integral can be approximated to be
$-|\psi_{0,0}\psi_{1,1}^*|\sin(kx+kz+\varphi_{1,1}
-\varphi_{0,0})=|\psi_{0,0}\psi_{1,1}^*|\cos(kx+kz+\varphi_{1,1}-\varphi_{0,0}
+\pi/2)$. Based on light diffraction theory, when a beam is incident
at an angle of $45^\circ$ onto a grating, which is the combination
of two gratings, the maximum diffraction light along the incident
direction occurs when these two gratings are in phase. That is,
$-\varphi_{-1,-1} + \varphi_{0,0} =\varphi_{1,1}
-\varphi_{0,0}+\pi/2$, or
\begin{equation}\label{phMat}
(\varphi_{1,1}-\varphi_{0,0})+\frac{\pi}{2}+(\varphi_{-1,-1}-\varphi_{0,0})=0.
\end{equation}
That means $\Delta\varphi_{0,1}=\pi/2$. According to our simulation
we find that  the relative phase of the matter waves $\Delta
\varphi_{0,1}$ is nearly $\pi/2$, and there is a little deviation
for different initial relative phases $\phi_0$ of pump beam. The
dependence of $\Delta \varphi_{0,1}$ on $ \phi_0$ reflects the
intrinsic nature of nonlinearity in the superradiance process since
it should not depend on the initial relative phase between the two
frequency components when the pumping pulse duration equals to an
exact cycle for a linear process. The initial relative phase of the
pumping beams slightly alters the phase matching condition, which
maybe results in the divergence between our simple model and the
numerical result. However, in our discussed model, the result giving
$\Delta\varphi_{0,1}=\pi/2$ is reasonable and meaningful for
understanding the scattering picture.

According to Eq.~({\ref{results}) and ({\ref{phMat}}) and neglecting $\alpha$, the ratio of backward scattered atoms to the total atom number versus the initial phase $\phi_{0}$ is simply a sinusoidal function and is plotted as the dotted line in Fig.~\ref{experdata}, with its amplitude adjusted to the experimental data. Backward scattering is enhanced when the relative phase is nearly $\phi_{0}=\pi/2$, while suppressed for $\phi_{0}$ close to $3\pi/2$. This simple model agrees well with our experimental data except for a small shift of the minimum.

Furthermore, the ratio of the backward scattering atom number to the total atom number is simulated numerically using coupled Maxwell-Schr\"{o}dinger equations~\cite{Zobay2006pra,yang} with an initial seed of one atom in modes $(1,1)$ and $(-1,-1)$, as shown by the dashed line in Fig.~\ref{experdata} with the coupling factor $g=1.5\times 10^{6}$. This simulation is in good qualitative agreement with our data but not completely satisfactory. The
difference may be due to the necessity to take into account higher order modes in the scattering process and to go beyond 1D approximation. The matching condition of matter waves may also not be exactly satisfied. Describing this mechanism under full quantum methods~\cite{guo} and considering wave mixing for the nonlinear high-order scattering  modes is our future work.

\begin{figure}[t]
   \begin{center}
       \includegraphics[height=6cm]{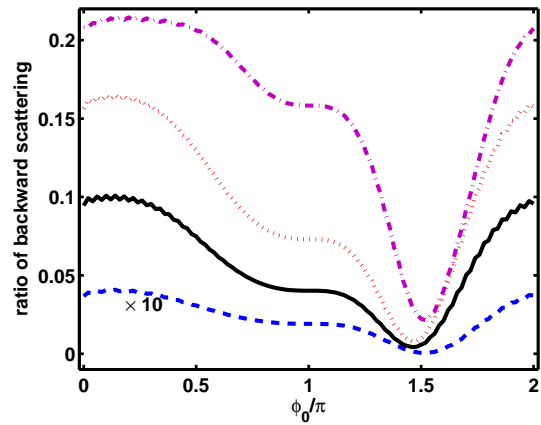}
       \end{center}
   \caption{The simulation for the ratio of backward scattered atoms to the total atom number vs the initial phase $\phi_{0}$ for different weak coupling factors $g$: $1.0\times10^{6}$ (dashed line, its value is amplified by 10 times to show); $1.5\times 10^{6}$(solid line); $1.6\times 10^{6}$(dotted line) and $1.7\times 10^{6}$(dashed-dotted line). }
   \label{gas}
\end{figure}

To show the features of the position of the maximum and minimum of the scattering ratio vs the relative phase $\phi_{0}$, we simulate it for different pumping factors $g$ in the week coupling regime, as shown in Fig.~\ref{gas}. The backward scattering ratio increases with the coupling factor, which depends on the system parameters like the pump beam intensity and detuning, the total number of atoms or the BEC shape. The simulation clearly shows that the position of the maximum and minimum of the scattering very weakly depends on the experimental parameters, this is a generic feature. However, the
value of the maximum increases with the coupling factor.

\begin{figure}[t]
   \begin{center}
   \includegraphics[height=6cm]{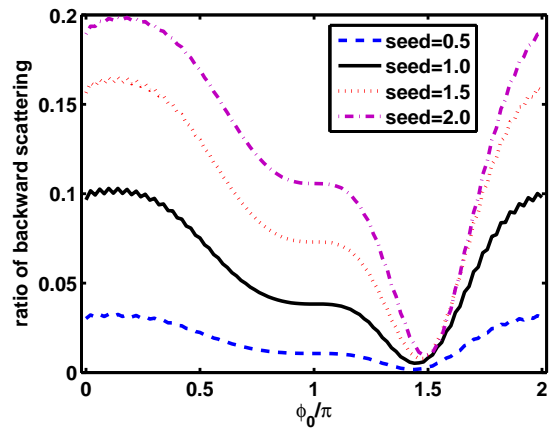}
   \end{center}
   \caption{The ratio of backward scattered atoms to the total atom number vs the initial phase
  $\phi_{0}$ between the two pump beams for different seeds with $g=1.5\times 10^{6}$.}
\label{seed}
\end{figure}

The superradiant scattering is initiated by quantum mechanical
noise, i.e., spontaneous Rayleigh scattering from individual
condensate atoms. Subsequent stimulated scattering and bosonic
enhancement lead to rapid growth of the side-mode populations. The
ratio of backward scattered atoms to the total atom number vs the
initial phase for different initial seeds in modes $(1,1)$ and
$(-1,-1)$ are shown in Fig.~\ref{seed}. We can see that the
positions of maximum and minimum do not vary with different seeds,
although the amplitude of the scattering events changes obviously.
The smaller the seeds, the lower the backward scattering amplitude.
This shows that these features caused by the initial relative phase
are not related to quantum fluctuation effects.

The above discussion is for the condition $\Delta\omega t=2\pi$,
which means identical coupling factor between light and atoms for
the different initial relative phase. If the condition $\Delta\omega
t=2\pi$ is violated, the integration of the pump intensity is
different for different relative phases, so that the coupling factor
is different. The ratio of backward scattered atoms to the total
atom number vs the initial phase $\phi_{0}$ is shown in Fig.7 for
the same frequency difference $15 \mathrm{KHz}$ and different pump
times $t$. From Fig.~\ref{violate}, we know that the position of the
extrema will change at different $t$, while the value of the maximum
will increase with $t$.

\begin{figure}
  \begin{center}
    \includegraphics[height=6cm]{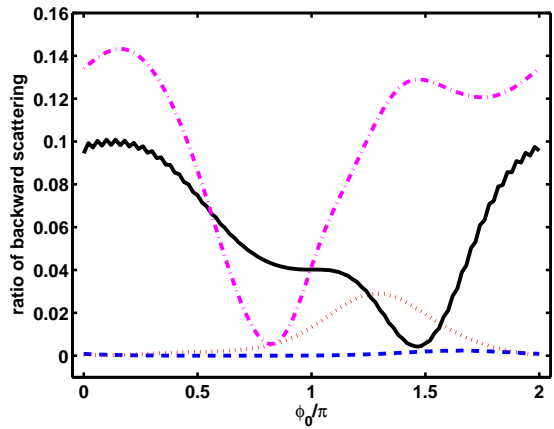}
  \end{center}
  \caption{The ratio of backward scattered atoms to the total atom number vs the initial phase $\phi_{0}$ for the same frequency difference $15$KHz but different pump times $t$: 22.22 $\mu$s (dashed line); 44.44 $\mu$s (dotted line); 66.67 $\mu$s (solid line) and 88.89 $\mu$s (dashed-dotted line). $g=1.5\times 10^{6}$.}
  \label{violate}
\end{figure}

In summary, we have studied superradiant scattering from a
Bose-Einstein condensate with a two-frequency pump beam. We have
shown that this phenomenon depends much on the initial relative
phase between the two optical components. Two matter gratings, one
formed by the condensate at rest plus the side mode with positive
momentum, the other by the condensate at rest plus the side mode
with negative momentum, scatter endfire modes photons, which in the
end affects the atom scattering process. We found that the relative
optical phase is imprinted on matter wave gratings and can enhance
or annihilate scattering in backward modes. Adjusting this phase
provides a powerful tool for controlling superradiant scattering.
It is already very beneficial for understanding the interaction
between matter waves and optical waves in cooperative scattering.

We thank the anonymous referee for useful and detailed suggestions.
This work is  partially supported by the state Key Development
Program for Basic Research of China (No.2005CB724503, 2006CB921401,
2006CB921402),NSFC (No.10874008, 10934010 and 60490280) and the
Project-sponsored by SRF for ROCS, SEM.

\end{document}